\documentclass{ws-mpla}

\newcommand{\He}{{}^3\mathrm{He}}

\newcommand{\Hh}{{}^3\mathrm{H}}
\newcommand{\pHe}{p\text{-}{}^3\mathrm{He}}

\newcommand{\mbf}[1]{{\mathbf{#1}}}

\newcommand{\cm}{\mathrm{c\!\:\!.m\!\:\!.}}

\begin{document}

\markboth{A. Deltuva}
{Faddeev-type calculations of few-body nuclear reactions
including Coulomb interaction}

\catchline{}{}{}{}{}

\title{FADDEEV-TYPE CALCULATIONS OF FEW-BODY NUCLEAR REACTIONS 
INCLUDING COULOMB INTERACTION}

\author{\footnotesize A. DELTUVA}

\address{Centro de F\'{\i}sica Nuclear da Universidade de Lisboa,
P-1649-003 Lisboa, Portugal}

\maketitle

\pub{Received (Day Month Year)}{Revised (Day Month Year)}

\begin{abstract}
The method of screening and renormalization is used to include
the Coulomb interaction between the charged particles in the 
description of few-body nuclear reactions. 
Calculations are done in the framework of Faddeev-type equations in
momentum-space.
The reliability of the method is demonstrated.
The Coulomb effect on observables is discussed.
\end{abstract}

\ccode{PACS Nos.: 21.45.+v, 21.30.-x, 24.70.+s, 25.10.+s}

\section{Introduction \label{sec:intro}}

The inclusion of the Coulomb interaction in the description of the
three-particle scattering is a challenging task in theoretical few-body 
nuclear physics. Due to its long range, the Coulomb potential $w(r)$ does
not satisfy the mathematical properties required for the formulation
of the standard scattering theory.
There is a number of suggestions how to overcome this difficulty;
most of them are based on the configuration-space 
framework\cite{kievsky:96a,chen:01a,ishikawa:03a,doleschall:05a} and are 
limited to energies below three-body breakup threshold (3BBT), while the 
others\cite{alt:04a,kadyrov:05a,oryu:06a}  have not matured yet
into practical applications. Up to now only few approaches led
to the results above 3BBT. Those are configuration-space calculations
for proton-deuteron ($p$-$d$) elastic scattering 
using the Kohn variational principle\cite{kievsky:01a} and
the screening and renormalization approach in the framework of momentum-space 
integral equations;\cite{alt:94a,alt:02a,deltuva:05a,deltuva:05d} 
the latter method will be discussed in more details.
Very recently $p$-$d$ results above 3BBT were also obtained using
modified Faddeev equation in configuration space together with the dumping
(screening) of particular Coulomb contributions.\cite{ishikawa:apfb08}

\section{Method of screening and renormalization \label{sec:th} } 

Our treatment of the Coulomb interaction is based on the idea of
screening and renormalization proposed in Refs.~\refcite{taylor:74a,semon:75a}
for the scattering of two charged particles.
The standard scattering theory is formally applicable to the screened 
Coulomb potential which, in the $r$-space representation, we choose as
\begin{equation} \label{eq:wr}
w_R(r) = w(r)\; e^{-(r/R)^n},
\end{equation}
where $R$ is the screening radius,
and $n$ controls the smoothness of the screening.
In 1974 Taylor\cite{taylor:74a} suggested that even for the description
of systems with the Coulomb interaction the standard scattering theory
may be useful, since in nature the Coulomb potential is
always screened. The study of the two-particle system with the screened
Coulomb interaction revealed that, as expected, the physical observables
become insensitive to screening provided it takes place at sufficiently
large distances $R$ and, in the $R \to \infty$ limit,
coincide with the corresponding quantities known from the analytical solution
of the two-particle Coulomb problem: though
the  on-shell screened Coulomb transition matrix 
$\langle \mbf{p}_f| t_R (e_i+i0) |\mbf{p}_i \rangle$ with  energy $e_i$ and
momenta $p_f=p_i$
 diverges in the $R \to \infty$ limit, after renormalization 
by (an equally) diverging phase factor $z_{R}^{-1}(p_i)$
it converges as a distribution to the well known proper Coulomb amplitude 
$\langle \mbf{p}_f| t_C |\mbf{p}_i \rangle$, i.e.,
  \begin{equation} \label{eq:tc}
    \lim_{R \to \infty}
    \langle \mbf{p}_f| t_R (e_i+i0) |\mbf{p}_i \rangle z_R^{-1}(p_i) 
    =  \langle \mbf{p}_f| t_C |\mbf{p}_i \rangle.
  \end{equation}
Renormalization by $ z_{R}^{-\frac12}(p_i)$ in the $R \to \infty$ limit relates 
also the screened and proper Coulomb wave functions.\cite{gorshkov:61}
The convergence of the scattering amplitude as a distribution corresponds 
to the physical conditions in a real experiment where the incoming beam 
is not a plane wave but a wave packet and the forward scattering cannot be 
measured.\cite{taylor:74a}
 This justifies the replacement (\ref{eq:tc}) in practical calculations.
We emphasize that the renormalization  (\ref{eq:tc}) relates
only the scattering amplitudes; its application to the off-shell transition
matrices\cite{oryu:08} in unjustified. However, for the calculation of 
observables, only the renormalization of the on-shell transition matrix and 
wave function is needed.

The screening and renormalization approach can be applied to more complicated 
systems.\cite{alt:80a}
 Here we briefly recall the procedure which is described in details in 
Refs.~\refcite{deltuva:05a,deltuva:05d}.
In the  multichannel on-shell transition matrix
$ \langle f| T^{(R)}_{\beta \alpha} (E_i+i0) |i \rangle $
between initial and final channel states $|i \rangle$ and $|f \rangle$,
$E_f = E_i$, derived from nuclear plus screened Coulomb potentials,
one has to isolate the diverging screened Coulomb contributions 
in the form of a two-body on-shell transition matrix
and two-body wave function with known renormalization properties.
This can be achieved using the two-potential formalism
 as long as in the initial/final states
there are no more than two charged bodies (clusters).
This also enables to decompose  
$ \langle f| T^{(R)}_{\beta \alpha} (E_i+i0) |i \rangle $ 
into contributions with different range properties. 
The long-range part, the  two-body on-shell transition matrix
$ \langle f| T^{c.m.}_{\alpha R} (E_i+i0) |i \rangle $,
derived from the screened  Coulomb potential of the form (\ref{eq:wr}) between
the centers of mass (c.m.) of the two charged bodies in the initial state,
is present in the elastic scattering only.
After renormalization by the diverging phase factor $Z_{iR}^{-1}$,
this contribution converges  towards its  $R \to \infty$ limit very slowly,
and, in general, as a distribution only, but the result 
$ \langle f| T^{c.m.}_{\alpha C}|i \rangle $, the pure Coulomb amplitude 
of two-body nature, is known analytically.
The remaining part of the elastic scattering amplitude
as well as the amplitudes for transfer and breakup are short-range 
operators that are externally distorted by Coulomb.
Due to their short-range nature, convergence with $R$ 
after the renormalization by the corresponding phase factors is fast
and, therefore,  the $R \to \infty$ limit can be calculated numerically
with high accuracy at finite $R$. Thus, the physical
scattering amplitudes are obtained after renormalization of
$ \langle f| T^{(R)}_{\beta \alpha} (E_i+i0) |i \rangle $ in the 
 $R \to \infty$ limit as
\begin{gather} \label{eq:UC}
\begin{split}
 \langle f| T_{\beta \alpha} |i \rangle = {} & 
 \delta_{\beta\alpha}  \langle f| T^{c.m.}_{\alpha C}|i \rangle \\ &
+\lim_{R \to \infty} \{ Z_{fR}^{-\frac12}
    \langle f | [ T^{(R)}_{\beta \alpha}(E_i + i0)   -
              \delta_{\beta\alpha} T^{\cm}_{\alpha R}(E_i + i0)]
            |i \rangle Z_{iR}^{-\frac12} \}.
\end{split}
\end{gather}

One can use standard scattering theory to calculate the 
multichannel transition operators  $T^{(R)}_{\beta \alpha} (E_i+i0)$ 
at finite screening radius $R$ and make sure that 
$R$ is large enough for the convergence of the results.
We solve  Alt, Grassberger, and Sandhas (AGS) equations
for three- and four-particle scattering\cite{alt:67a,grassberger:67}
which are equivalent to Faddeev and 
Yakubovsky equations.\cite{faddeev:60a,yakubovsky:67}
We employ momentum-space partial-wave representation as described 
in detail in Refs.~\refcite{chmielewski:03a,deltuva:03a,deltuva:07a}
for three- and four-nucleon scattering without the Coulomb force.
However, the screened Coulomb interaction,
due to its longer range, compared to the nuclear interaction, 
brings additional difficulties: quasisingular nature of the potential
and slow convergence of the partial-wave expansion.
The right choice of  the parameter $n$ controlling the smoothness 
of the screening (\ref{eq:wr}) is essential in resolving those difficulties
as we demonstrated in Ref.~\refcite{deltuva:05a}.
We  work with a sharper screening than the Yukawa screening
$(n=1)$ of Refs.~\refcite{alt:94a,alt:02a}. We want to ensure that the
screened Coulomb potential $w_R(r)$ approximates well the true Coulomb one
$w(r)$ for distances $r<R$  and simultaneously vanishes rapidly for $r>R$,
providing a comparatively fast convergence of the partial-wave expansion
and less pronounced quasisingularities.
In contrast, the sharp cutoff  $(n \to \infty)$
yields an unpleasant oscillatory behavior in the momentum-space representation,
leading to convergence problems. Depending on the reaction
we found the values $3 \le n \le 8$ to provide a sufficiently smooth,
but at the same time a sufficiently rapid screening around $r=R$.
In any case the screening radius $R$ needed for convergence in 
Eq.~\eqref{eq:UC} is considerably larger than the range of the nuclear 
interaction and, therefore, the calculation of $T^{(R)}_{\beta \alpha} (E_i+i0)$ 
requires the inclusion of partial waves with angular momentum much higher
than needed for the nuclear potential alone.
This problem can be solved in an efficient and reliable way either by using
the perturbative approach for high two-particle partial waves, developed in
Ref.~\refcite{deltuva:03b}, or even without it as discussed in 
Ref.~\refcite{deltuva:06a}.

The internal criterion for the reliability of our method is the convergence
of the observables with screening radius $R$ used to calculate the 
Coulomb-distorted short-range part of the amplitudes in Eq.~\eqref{eq:UC}.
Numerous examples for three-nucleon hadronic and electromagnetic reactions,
$\alpha$-$d$ and $\pHe$ scattering  can be found in 
Refs.~\refcite{deltuva:05a,deltuva:05d,deltuva:06b,deltuva:fb18,deltuva:07b}.
In most cases the convergence is impressively fast;
the screening radius $R = 10 $ to 30 fm is sufficient.
The exceptions requiring larger screening radii are the 
observables at very low energies and the breakup
differential cross section in kinematical situations characterized 
by very low relative energy $E_{\mathrm{rel}}$ between the two
charged particles, e.g., $p$-$d$ breakup or photodisintegration of $\He$
close to the $pp$ final-state interaction  ($pp$-FSI) regime.\cite{deltuva:05d}
In there, the Coulomb repulsion is responsible for decreasing the cross section,
converting the FSI peak obtained in the absence of Coulomb
into a minimum with zero cross section at  $E_{\mathrm{rel}}=0$.
Such a behavior is seen in the experimental data as 
well.\cite{kistryn:06a,sagara:fb18}
The slow convergence under those conditions is not surprising, since the 
renormalization factor itself as well as the Coulomb parameter
become ill-defined, indicating that the  screening and renormalization
procedure cannot be applied at $E_{\mathrm{rel}}=0$.
Therefore an extrapolation has to be used to calculate
the observables at $E_{\mathrm{rel}}=0$, which works pretty well since
the observables vary smoothly with $E_{\mathrm{rel}}$.

\begin{figure}[!]
\begin{center}
\includegraphics[scale=0.6]{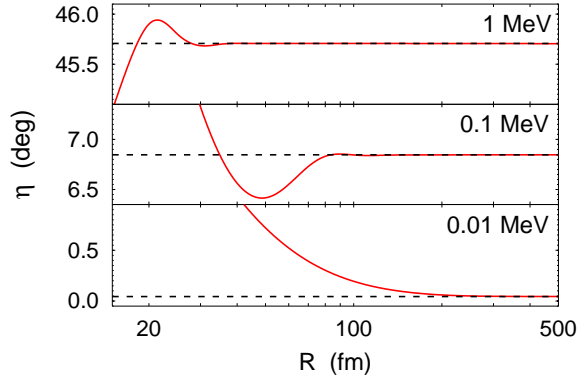}
\end{center} 
\caption{ 
${}^1S_0$ phase shifts of $pp$ scattering at 0.01, 0.1 and 1 MeV c.m. energy.
Results obtained using screening and 
renormalization method are shown as function of the screening radius 
$R$ (solid curves). Dashed lines are the exact results.
\protect\label{fig:pp}}
\end{figure}

It has been claimed\cite{oryu:08}
that the method of screening and renormalization ceases to work
even for $pp$ scattering below 0.1 MeV c.m. energy. In Fig.~\ref{fig:pp}
we prove that this is not so by getting well converged results at 0.01 MeV.
However,  Fig.~\ref{fig:pp} also shows that the 
screening radius indeed has to be increased with decreased energy
and at some point would become too large for reliable numerical calculation.

\section{Results and summary \label{sec:summ}}

We have discussed  how the Coulomb interaction between the charged
particles can be included into the description of few-body reactions
using the old idea of screening and renormalization.\cite{taylor:74a}
The calculations are done in the framework of AGS integral 
equations\cite{alt:67a,grassberger:67} in  momentum-space. 
The screening and renormalization approach has already been used
for $p$-$d$ scattering\cite{alt:94a,alt:02a,berthold:90}
but with limited success: those calculations 
were based on quasiparticle equations with 
rank-1 separable potentials and, in addition, the screened Coulomb 
transition matrix  was approximated by the screened Coulomb potential;
none of these approximations is used by us.
It is the new screening function that allows us to avoid these approximations
and obtain fully converged results.
Furthermore, the results for $p$-$d$ elastic scattering obtained by the 
present technique were compared\cite{deltuva:05b} with
those of Ref.~\refcite{kievsky:01a} obtained from the variational solution
of the three-nucleon Schr\"odinger equation in configuration space
with the inclusion of an \emph{unscreened} Coulomb potential
between the protons and imposing  the proper Coulomb boundary
conditions explicitly. Good agreement over a wide energy range
was found indicating that both techniques for including the Coulomb interaction
are reliable. At very low energies the coordinate-space treatments remain 
favored since there the method of screening and renormalization 
converges slowly and therefore becomes technically too demanding, 
but at higher energies and for three-body breakup reactions it is more 
efficient. 
A detailed comparison still has to be done but it seems that 
there is a reasonable agreement between momentum- and coordinate-space
results also  in  the case of $p$-$d$ breakup\cite{ishikawa:apfb08} 
and $p$-$\He$ scattering.\cite{viviani:01a}

The present method was used to study $p$-$d$ elastic scattering and  breakup in
Refs.~\refcite{deltuva:05a,deltuva:05d,kistryn:06a,sagara:fb18,ley:06},
 $p$-$d$ radiative capture in Refs.~\refcite{deltuva:05a,klechneva:06}
and photo- and electrodisintegration of $\He$ in 
Refs.~\refcite{deltuva:05a,deltuva:05d,sauer:fb18}.
Furthermore, it was applied to the nuclear reactions
dominated by three-body degrees of freedom, i.e., deuteron scattering
on stable nuclei or proton scattering on one-neutron halo nuclei.
Examples are low-energy $\alpha$-$d$ elastic scattering and 
breakup\cite{deltuva:06b}
and $d+{}^{12}\mathrm{C}$ and  $p+{}^{11}\mathrm{Be}$ elastic,
transfer, and breakup reactions\cite{deltuva:07d,crespo:07a,crespo:08a}
where also the accuracy of traditional approximate nuclear reaction 
approaches like Continuum Discretized Coupled Channels (CDCC) method, Glauber,
and Distorted Wave Impulse Approximation (DWIA) could be tested.
Finally, all elastic and transfer four-nucleon reactions below
three-body breakup threshold, i.e.,
$n+\Hh$, $p+\He$, $p+\Hh$, $n+\He$, and $d+d$,  have been studied in 
Refs.~\refcite{deltuva:07a,deltuva:07b,deltuva:07c,deltuva:08a}.
The conclusion is that the Coulomb effect  is 
important at low energies for all kinematic regimes, but gets confined 
to the forward direction in elastic scattering at higher energies.
In $p$-$d$ breakup and in three-body e.m. disintegration of $\He$ 
the Coulomb effect is extremely important in kinematical regimes close to 
$pp$-FSI. There the $pp$ repulsion converts the $pp$-FSI peak obtained in the
absence of Coulomb into a minimum with zero cross section.\cite{sagara:fb18}
This significant change of the cross section behavior has important
consequences in nearby configurations where one may observe
instead an increase of the cross section due to Coulomb.\cite{kistryn:06a}
However, some of the long-standing discrepancies between experiment and theory
like the space star anomaly in $p$-$d$ breakup are not resolved by the inclusion
of the Coulomb interaction.\cite{sagara:fb18,deltuva:05c}
A very strong Coulomb effect is found in $\alpha$-$d$ breakup
where the shift of $\alpha p$ $P$-wave resonance
position leads to the corresponding shifts of the differential cross
section peaks.\cite{deltuva:06b}

\section*{Acknowledgments}
This work has been performed in collaboration with A.~C.~Fonseca and 
P.~U.~Sauer. The author is supported by the Funda\c{c}\~{a}o para a 
Ci\^{e}ncia e a Tecnologia (FCT) grant SFRH/BPD/34628/2007.



\begin{thebibliography}{10}

\bibitem{kievsky:96a}
A. Kievsky  {\it et~al.}, {\it Nucl. Phys.} {\bf A607}, 402 (1996).

\bibitem{chen:01a}
C.~R. Chen, J.~L. Friar, and G.~L. Payne, {\it Few-Body Syst.} {\bf 31},  13
  (2001).

\bibitem{ishikawa:03a}
S. Ishikawa, {\it Few-Body Syst.} {\bf 32},  229  (2003).

\bibitem{doleschall:05a}
P. Doleschall and Z. Papp, {\it Phys.~Rev.~C} {\bf 72},  044003  (2005).

\bibitem{alt:04a}
E.~O. Alt, S.~B. Levin, and S.~L. Yakovlev, {\it Phys.~Rev.~C} {\bf 69},
  034002  (2004).

\bibitem{kadyrov:05a}
A.~S. Kadyrov  {\it et~al.}, {\it  Phys.~Rev.~A} {\bf 72},  032712  (2005).

\bibitem{oryu:06a}
S. Oryu, {\it Phys.~Rev.~C} {\bf 73},  054001  (2006).

\bibitem{kievsky:01a}
A. Kievsky, M. Viviani, and S. Rosati, {\it Phys.~Rev.~C} {\bf 64},  024002
  (2001).

\bibitem{alt:94a}
E.~O. Alt and M. Rauh, {\it Few-Body Syst.} {\bf 17},  121  (1994).

\bibitem{alt:02a}
E.~O. Alt {\it et~al.}, {\it  Phys.~Rev.~C} {\bf 65},  064613  (2002).

\bibitem{deltuva:05a}
A. Deltuva, A.~C. Fonseca, and P.~U. Sauer, {\it Phys.~Rev.~C} {\bf 71},
  054005  (2005).

\bibitem{deltuva:05d}
A. Deltuva, A.~C. Fonseca, and P.~U. Sauer, {\it Phys.~Rev.~C} {\bf 72},
  054004  (2005).

\bibitem{ishikawa:apfb08}
S. Ishikawa, talk at this conference.

\bibitem{taylor:74a}
J.~R. Taylor, {\it Nuovo Cimento} {\bf B23},  313  (1974).

\bibitem{semon:75a}
M.~D. Semon and J.~R. Taylor, {\it Nuovo Cimento} {\bf A26},  48  (1975).

\bibitem{gorshkov:61}
V.~G. Gorshkov, {\it Sov.~Phys.-JETP} {\bf 13},  1037  (1961).

\bibitem{oryu:08}
S. Oryu,  talk at this conference.

\bibitem{alt:80a}
E.~O. Alt and W. Sandhas, {\it Phys.~Rev.~C} {\bf 21},  1733  (1980).

\bibitem{alt:67a}
E.~O. Alt, P. Grassberger, and W. Sandhas, {\it Nucl.~Phys.} {\bf B2},  167
  (1967).

\bibitem{grassberger:67}
P. Grassberger and W. Sandhas, {\it Nucl. Phys.} {\bf B2},  181  (1967); 
E. O. Alt, P. Grassberger, and W. Sandhas, JINR report No. E4-6688 (1972).

\bibitem{faddeev:60a}
L.~D. Faddeev, {\it Zh.~Eksp.~Theor.~Fiz.} {\bf 39},  1459  (1960) 
[{\it Sov.~Phys.  JETP} {\bf 12}, 1014 (1961)].

\bibitem{yakubovsky:67}
O.~A. Yakubovsky, {\it Yad. Fiz.} {\bf 5},  1312  (1967) 
[{\it Sov. J. Nucl. Phys.}  {\bf 5}, 937 (1967)].

\bibitem{chmielewski:03a}
K. Chmielewski {\it et~al.}, {\it Phys.~Rev.~C} {\bf 67},  014002  (2003).

\bibitem{deltuva:03a}
A. Deltuva, K. Chmielewski, and P.~U. Sauer, {\it Phys.~Rev.~C} {\bf 67},
  034001  (2003).

\bibitem{deltuva:07a}
A. Deltuva and A.~C. Fonseca, {\it Phys.~Rev.~C} {\bf 75},  014005  (2007).

\bibitem{deltuva:03b}
A. Deltuva, K. Chmielewski, and P.~U. Sauer, {\it Phys.~Rev.~C} {\bf 67},
  054004  (2003).

\bibitem{deltuva:06a}
A. Deltuva, A.~C. Fonseca, and P.~U. Sauer, {\it Phys.~Rev.~C} {\bf 73},
  057001  (2006).

\bibitem{deltuva:06b}
A. Deltuva, {\it Phys.~Rev.~C} {\bf 74},  064001  (2006).

\bibitem{deltuva:fb18}
A. Deltuva, A.~C. Fonseca, and P.~U. Sauer, {\it Nucl.~Phys.} {\bf A790},  52c
  (2007).

\bibitem{deltuva:07b}
A. Deltuva and A.~C. Fonseca, {\it Phys.~Rev.~Lett.} {\bf 98},  162502  (2007).

\bibitem{kistryn:06a}
S. Kistryn~{\it et al.}, {\it Phys.~Lett.~B} {\bf 641},  23  (2006).

\bibitem{sagara:fb18}
K. Sagara~{\it et al.}, {\it Nucl.~Phys.} {\bf A790},  348c  (2007).

\bibitem{berthold:90}
G.~H. Berthold, A. Stadler, H. Zankel, {\it Phys.~Rev.~C} {\bf 41}, 1365 (1990).

\bibitem{deltuva:05b}
A. Deltuva {\it et~al.}, {\it Phys.~Rev.~C} {\bf 71},  064003  (2005).

\bibitem{viviani:01a}
M. Viviani  {\it et~al.}, {\it Phys.~Rev. Lett.} {\bf 86}, 3739  (2001).

\bibitem{ley:06}
J. Ley  {\it et~al.}, {\it Phys.~Rev.~C} {\bf 73},  064001  (2006).

\bibitem{klechneva:06}
T. Klechneva  {\it et~al.}, {\it Phys.~Rev.~C} {\bf 73},  034005  (2006).

\bibitem{sauer:fb18}
A. Deltuva, A.~C. Fonseca, and P.~U. Sauer, {\it Nucl.~Phys.} {\bf A790},  344c
   (2007).

\bibitem{deltuva:07d}
A. Deltuva {\it et~al.}, {\it Phys.~Rev.~C} {\bf 76},  064602  (2007).

\bibitem{crespo:07a}
R. Crespo {\it et~al.}, {\it Phys.~Rev.~C} {\bf 76},  014620  (2007).

\bibitem{crespo:08a}
R. Crespo {\it et~al.}, {\it Phys.~Rev.~C} {\bf 77},  024601  (2008).

\bibitem{deltuva:07c}
A. Deltuva and A.~C. Fonseca, {\it Phys.~Rev.~C} {\bf 76},  021001(R)  (2007).

\bibitem{deltuva:08a}
A. Deltuva, A.~C. Fonseca, and P.~U. Sauer, {\it Phys.~Lett.~B} {\bf 660},  471
   (2008).

\bibitem{deltuva:05c}
A. Deltuva, A.~C. Fonseca, and P.~U. Sauer, {\it Phys.~Rev.~Lett.} {\bf 95},
  092301  (2005).

\end{thebibliography}

\end{document}